\documentclass[aps,prb,twocolumn,superscriptaddress,showpacs,floatfix]{revtex4}%
\usepackage{graphicx}%

\begin{document}

\title{Search for quantum dimer phases and transitions in a frustrated spin ladder}

\author{Hsiang-Hsuan Hung}
\affiliation{National Key Laboratory of Solid State Microstructure
and Department of Physics, Nanjing University, Nanjing 210093}

\author{Chang-De Gong}
\affiliation{Chinese Center of Advanced Science and Technology
(World Laboratory) P.O. Box 8730, Beijing 100080}
\affiliation{National Key Laboratory of Solid State Microstructure
and Department of Physics, Nanjing University, Nanjing 210093}

\author{Yung-Chung Chen}
\affiliation{Department of Physics, Tunghai University, Taichung}

\author{Min-Fong Yang}
\affiliation{Department of Physics, Tunghai University, Taichung}

\begin{abstract}
A two-leg spin-1/2 ladder with diagonal interactions is
investigated numerically. We focus our attention on the
possibility of columnar dimer phase, which was recently predicted
based on a reformulated bosonization theory. By using density
matrix renormalization group technique and exact diagonalization
method, we calculate columnar dimer order parameter, spin
correlation on a rung, string order parameters, and scaled
excitation gaps. Carefully using various finite-size scaling
techniques, our results show no support for the existence of
columnar dimer phase in the spin ladder under consideration.
\end{abstract}


\maketitle


Much effort has been devoted to understanding the effects of
competing interactions on quasi-one-dimensional systems. The
possibility of unconventionally ordered phases has been the focus
of interest. For example, one-dimensional extended Hubbard model
(EHM) with nearest neighbour repulsion $V$ in addition to on-site
repulsion $U$ has been investigated
intensively.~\cite{Emery-Solyom,Hirsch,Cannon,Voit,Dongen,Jeckelmann,Nakamura,TF,Sengupta,Sandvik,Zhang,Tam}
It had been considered for a long time that the ground-state phase
diagram at half filling has only two phases, the spin-density-wave
(SDW) and the charge-density-wave (CDW)
states.~\cite{Emery-Solyom,Hirsch,Cannon,Voit,Dongen,Jeckelmann}
Moreover, the order of the phase transition at $U \simeq 2V$ can
change from continuous to first order at a tricritical point,
which was speculated to exist in the intermediate coupling
regime.~\cite{Hirsch,Cannon,Voit} Later, by using the
level-spectroscopy method, Nakamura pointed out that there exists
also a novel spontaneously dimerized phase, the so-called
bond-order-wave (BOW) phase, in a narrow strip between the SDW and
the CDW phases in the weak coupling region.~\cite{Nakamura} While
this phase is absent in standard one-loop g-ology and bosonization
calculations,~\cite{Emery-Solyom,Voit} its existence was supported
by a reformulated one,~\cite{TF} where higher-order terms were
included. The appearance of the BOW phase was subsequently
confirmed by quantum Monte Carlo
simulations,~\cite{Sengupta,Sandvik} density matrix
renormalization group (DMRG) method,~\cite{Zhang} and functional
renormalization group calculations.~\cite{Tam}

Interestingly, same story may happen also in an antiferromagnetic
two-leg spin ladder with diagonal frustrations. The Hamiltonian of
this model reads as follows,
\begin{eqnarray}\label{ham}
H &=& J \sum_{\alpha=1,2} \sum_i {\bf S}_{\alpha,i} \cdot {\bf
S}_{\alpha,i + 1}
+ J_{\perp} \sum_i {\bf S}_{1,i} \cdot {\bf S}_{2,i} \nonumber\\
&&+ J_{\times} \sum_i \left(  {\bf S}_{1,i} \cdot {\bf S}_{2,i +
1} + {\bf S}_{1,i + 1} \cdot {\bf S}_{2,i} \right)  ,
\end{eqnarray}
where ${\bf S}_{\alpha,i}$ denotes a spin-$1/2$ operator at site
$i$ of the $\alpha$-th leg. $J$ and $J_{\perp}$ are the exchange
couplings on legs and rungs, respectively. We set $J\equiv 1$
hereafter. $J_{\times}$ denotes the next-nearest-neighbor
interchain coupling. This model has been investigated for
decades,~\cite{Gelfand,AEN,KFSS,FLS,Nakamura03,WEI,Wang,Nakamura-un}
and considerable amount of knowledge has been accumulated. It has
been believed that the ground state phase diagram consists of only
two phases, the rung-singlet (RS) phase and the so-called Haldane
phase. The earlier bosonization study predicted a direct
transition between these two phases at
$J_{\perp}=2J_{\times}$.~\cite{AEN} Various numerical
calculations, such as series expansions,~\cite{WEI}
DMRG,~\cite{Wang} and level-spectroscopy
method,~\cite{Nakamura-un} showed the phase boundary to be shifted
away from the $J_{\perp}=2J_{\times}$ line, with a larger RS
phase. Moreover, the phase transition seems to be of second order
at weak interchain couplings and becomes first order at stronger
couplings.~\cite{WEI,Wang} Recently, it was suggested that there
exists an intermediate, spontaneously dimerized phase, the
so-called columnar dimer (DC) phase, lying between the Haldane and
the RS phases at weak interchain couplings.~\cite{Oleg} This
remarkable proposal is based on a reformulated weak-coupling field
theory, which is similar in spirit to that in Ref.~\onlinecite{TF}
with success in the study of one-dimensional EHM. According to
Ref.~\onlinecite{Oleg}, for a given $J_{\times}$, the DC phase
occurs within a narrow but finite region $(J_{\perp})_{c,T} \leq
J_{\perp} \leq (J_{\perp})_{c,S}$ in the phase diagram. Here
$(J_{\perp})_{c,T}=2J_{\times}-5J_{\times}^{2}/\pi^{2}$ and
$(J_{\perp})_{c,S}=2J_{\times}-J_{\times}^{2}/\pi^{2}$ are two
distinct critical points, given by vanishing mass gaps in the
spin-triplet and the spin-singlet sectors, repsectively. As
mentioned before, this intermediate DC phase was not found in
previous numerical calculations.~\cite{WEI,Wang,Nakamura-un} Thus
the existence of this spontaneously dimerized DC phase is
surprising and calls for thorough theoretical studies.

In this paper, we try to find numerical evidence for the
possibility of the DC phase in a two-leg spin ladder of
Eq.~(\ref{ham}) in the vicinity of $J_{\perp}= 2J_{\times}$. Here
we take $J_{\times}=0.2$.~\cite{note} It is smaller than the value
$J_{\times}=0.287$, where the transition changes to be first
order.~\cite{Wang} Thus the proposed region for the DC phase
becomes $0.38 \leq J_{\perp} \leq 0.396$. The values of
$J_{\times}$ and $J_{\perp}$ under consideration should be small
enough in accordance with the weak-coupling field theory in
Ref.~\onlinecite{Oleg}. Both the DMRG technique~\cite{DMRG,Hung}
under open boundary conditions up to $L=400$ rungs and the
L\'anczos exact diagonalization method with periodic boundary
conditions up to $L=16$ are used. In our DMRG calculations, $500$
states per block are kept, and the truncation error is of the
order of $10^{-7}$. To demonstrate the possibility of the DC
phase, the most direct way is to show the corresponding order
parameter being nonzero in the proposed region of the phase
diagram. As far as we know, this order parameter has not yet been
measured for the present model. We note that the critical point
$(J_{\perp})_{c,T}$ in the spin-triplet sector is consistent with
the phase boundary found previously.~\cite{WEI,Wang,Nakamura-un}
Therefore, another support of the proposal in
Ref.~\onlinecite{Oleg} is to show the existence of another
critical point at $(J_{\perp})_{c,S}$ with vanishing mass gap in
the spin-singlet sector. Carefully using various finite-size
scaling skills for the DC order parameter and other physical
quantities defined below, our results fail to show the existence
of the DC phase, but instead they indicate a direct transition
between the RS and the Haldane phases.


\begin{figure}
\includegraphics[width=3.4in]{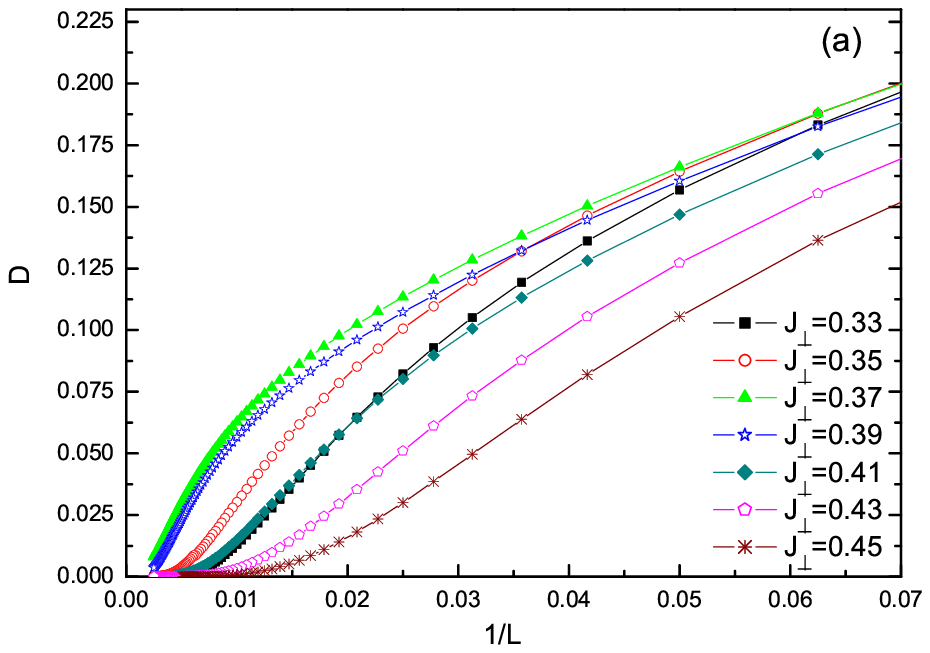}
\vskip -0.5cm \includegraphics[width=3.4in]{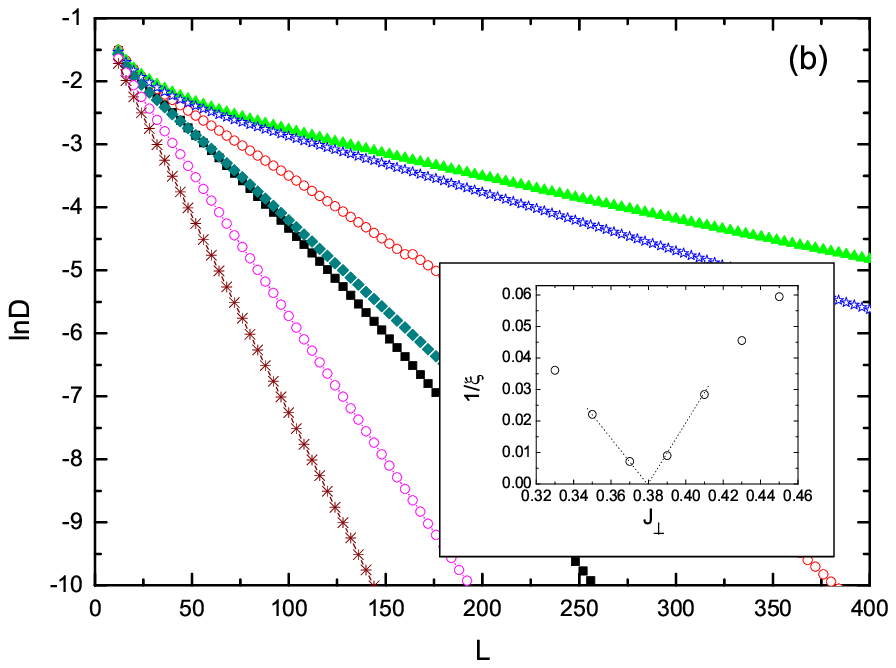}%
\vskip -0.5cm \caption{(Color online) (a) Size dependence of the
DC order parameter $D$ for various $J_\perp$ with $J_\times =0.2$.
(b) $\ln D$ versus $L$ for various values of $J_\perp$ with
$J_\times =0.2$. Labels for various $J_\perp$ are the same as
those in (a). Inset shows the inverse of the correlation length
$\xi$ for various values of $J_\perp$. } \label{fig:dimer}
\end{figure}

By using the DMRG technique under open boundary conditions, we
first analyze the DC order parameter.
As shown in Ref.~\onlinecite{Lecheminant}, due to the presence of
open ends, weak dimerization profiles can be induced near the
boundaries.
In order to reduce the boundary effect, the DC order parameter is
calculated by the difference of local spin correlation on legs,
\begin{equation}\label{eq:dimer2}
D \equiv \left| \left\langle \frac{1}{2}\sum_{\alpha=1,2} \left(
{\bf S}_{\alpha,i}\cdot {\bf S}_{\alpha,i+1} - {\bf
S}_{\alpha,i+1} \cdot {\bf S}_{\alpha,i+2} \right) \right\rangle
\right| \;,
\end{equation}
where only the bonds in the middle of finite open ladder with
length $L$ are considered (i.e., $i=L/2$). $\langle\cdots\rangle$
means the ground-state expectation value. The DC order parameter
in the thermodynamic limit is then $D_\infty \equiv
\lim_{L\to\infty} D$. Our results of the DC order parameter $D$
for various $J_\perp$ with $J_\times =0.2$ are shown in
Fig.~\ref{fig:dimer}~(a). We find that $D$ always decreases to
zero in the thermodynamic limit even for $J_{\perp}=0.39$, which
lies within the suggested region for the DC phase.
Fig.~\ref{fig:dimer}~(b) shows $\ln D$ versus $L$ for various
values of $J_\perp$. It is found that, for $L$ being large, $D
\approx c \exp (-L/\xi)$, where $c$ is a constant and $\xi$ is a
kind of correlation lengths. Moreover, as shown in the inset of
Fig.~\ref{fig:dimer}~(b), $1/\xi \propto
|J_{\perp}-(J_{\perp})_{c}|$ with $(J_{\perp})_{c}\approx 0.38$,
which agrees with the value of the proposed critical point
$(J_{\perp})_{c,T}=0.38$ in the spin-triplet sector. This
indicates that the long-range DC phase may appear only at this
phase boundary, rather than for a finite region in the phase
diagram. Besides, it shows no evidence for the additional
second-order quantum phase transition at
$J_{\perp}=(J_{\perp})_{c,S}=0.396$, since the correlation length
$\xi$ diverges only at a single critical point
$(J_{\perp})_{c,T}$, rather than at two points.


\begin{figure}
\includegraphics[width=3.4in]{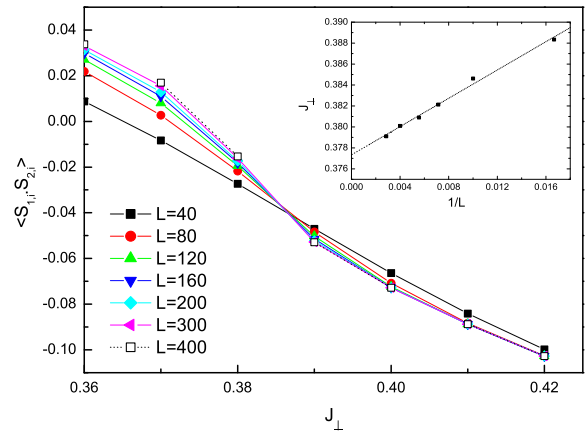}
\vskip -0.5cm \caption{(Color online) Spin correlation on the
($N$/2)-th rung as a function of $J_\perp$ for various sizes $L$
with $J_\times =0.2$. Inset shows the $L^{-1}$ scaling behavior of
the crossing points.}\label{fig:rung}
\end{figure}

To be sure if we miss the critical point $(J_{\perp})_{c,S}$ in
the above analysis, a finite-size crossing method~\cite{Venuti} is
used, which is applicable to detect the critical points of
second-order quantum phase transitions. It is noted that one can
always decompose the Hamiltonian into two parts, i.e.,
$H\equiv\mathcal{H}_{0}+g\mathcal{V}$, and consider the transition
being driven by the parameter $g$. Based on the finite-size
scaling analysis, it is shown that the curves of the mean value
$\mathcal{O}\equiv\langle\mathcal{V}\rangle/L$ at two successive
values of size $L$ as a function of $g$ will cross at a single
point $g_{L}^{\ast}$ near each critical point
$g_{c}$.~\cite{Venuti} The value of the critical point $g_{c}$ can
be found numerically by extrapolating the sequence $g_{L}^{\ast}$
to $L \rightarrow \infty$. In the present case, we take the
driving parameter as $J_\perp$. Thus the corresponding
transition-driving term becomes $\mathcal{V}=\sum_i {\bf S}_{1,i}
\cdot {\bf S}_{2,i}$, which gives $\mathcal{O} = \langle {\bf
S}_{1,i} \cdot {\bf S}_{2,i}\rangle$ by translational invariance
if periodic boundary conditions are used. In case of finite open
ladders, to avoid boundary effects, sites in the middle of ladders
(i.e., $i=L/2$) are used. In Fig.~\ref{fig:rung}, we plot the
curves of $\langle {\bf S}_{1,i} \cdot {\bf S}_{2,i}\rangle$
versus $J_\perp$ for various sizes $L$, which are calculated by
the DMRG technique. It is found that there is only one crossing
point $J_\perp^{\ast} (L)$ at $L \equiv (L_1 + L_2)/2$ for the
curves at two subsequent sizes $L_1$ and $L_2$. This indicates
that there is {\it only one}, but not two, phase transition. Our
conclusion is consistent with the picture provided by previous
investigations,\cite{Gelfand,AEN,KFSS,FLS,Nakamura03,WEI,Wang,Nakamura-un}
where the complete phase diagram consists of only two phases with
a single phase boundary. The scaling behavior of the crossing
points is shown in the inset of Fig.~\ref{fig:rung}. It is found
that the crossing points converge to the value
$(J_\perp)_{c}\simeq 0.378$ for the critical point. Our finding
agrees well with that obtained by previous DMRG
calculation~\cite{Wang} and is consistent with the predicted value
for $(J_{\perp})_{c,T}$.~\cite{Oleg} Again, our results provide no
support for the existence of another critical point at
$(J_{\perp})_{c,S}$ in the spin-singlet sector.


\begin{figure}
\includegraphics[width=3.4in]{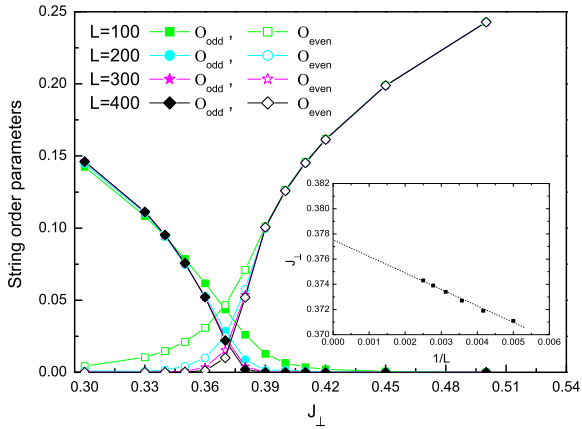}
\vskip -0.5cm \caption{(Color online) String order parameters for
ladders with various sizes $L$ as functions of $J_\perp$ with
$J_\times =0.2$. Inset shows the $L^{-1}$ scaling behavior of the
crossing points.} \label{fig:string}
\end{figure}

While our results show that there is no DC phase and there are
only two phases in the complete phase diagram, the nature of these
two phases has not yet been explored in the present study.
According to previous investigations, these two phases should be
the Haldane and the RS phases, and they can be identified by two
distinct string order parameters ${\cal O}_{\rm odd}$ and ${\cal
O}_{\rm even}$.~\cite{KFSS,FLS,Nakamura03,comment} These two
string order parameters are given by
\begin{equation}
{\cal O}_{P} =- \lim_{|i-j|\to\infty}\left\langle
\tilde{S}_{P,i}^z \exp \left( i\pi\sum_{l=i+1}^{j-1}
\tilde{S}_{P,l}^z \right) \tilde{S}_{P,j}^z \right\rangle,
\label{eq:string}
\end{equation}
where $P=$ odd, even. The composite spin operators are defined as
$\tilde{S}_{{\rm odd},i}^z = S_{1,i}^z + S_{2,i}^z$ and
$\tilde{S}_{{\rm even},i}^z = S_{1,i}^z + S_{2,i+1}^z$. Because of
spin isotropy, we calculate the string order parameters for the
$z$-component spins only. In case of finite ladders, it turns out
that the intersection of the curves of two distinct string order
parameters implies the critical point.~\cite{Nakamura03} In order
to reduce the undesirable boundary effects, we fix site $j$ in
Eq.~(\ref{eq:string}) at the center of the chain and let site
$i=20$ in our calculations. Our DMRG results of ${\cal O}_{\rm
odd}$ and ${\cal O}_{\rm even}$ for various sizes $L$ as functions
of $J_\perp$ with $J_{\times}=0.2$ are shown in
Fig.~\ref{fig:string}. For smaller $J_\perp$, one has ${\cal
O}_{\rm odd}\neq 0$ and ${\cal O}_{\rm even}=0$, which implies the
Haldane phase; while ${\cal O}_{\rm odd}=0$ and ${\cal O}_{\rm
even}\neq 0$ for larger $J_\perp$, which implies the RS phase. The
finite size scaling procedure is used to determine the
thermodynamic limit of the value of the critical point. As shown
in the inset of Fig.~\ref{fig:string}, the crossing points
converge to the value of $(J_\perp)_{c}\simeq 0.378$, which agrees
quite well with that found in the inset of Fig.~\ref{fig:rung}.


\begin{figure}
\includegraphics[width=3.4in]{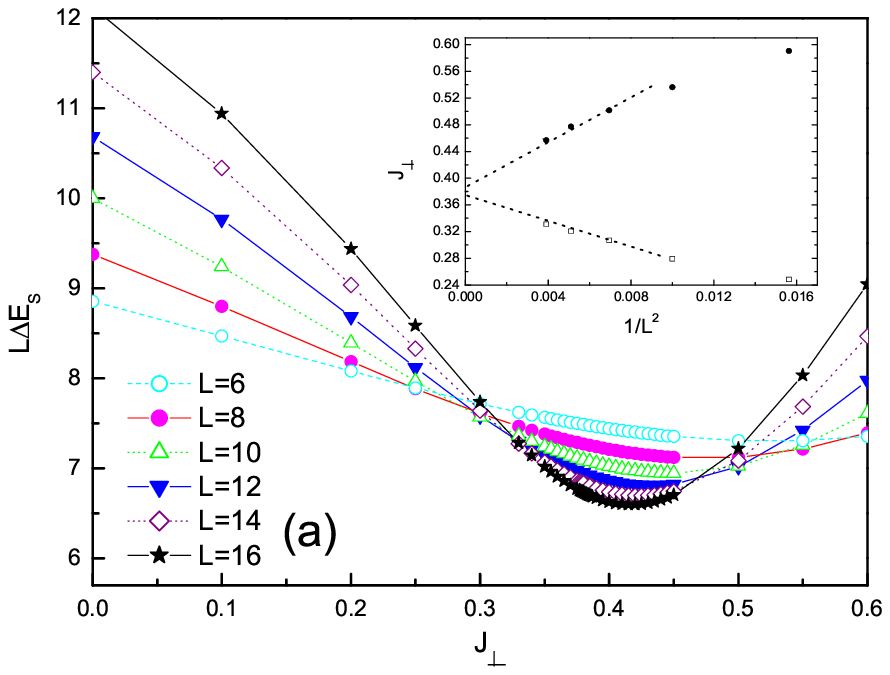}
\vskip -0.5cm \includegraphics[width=3.4in]{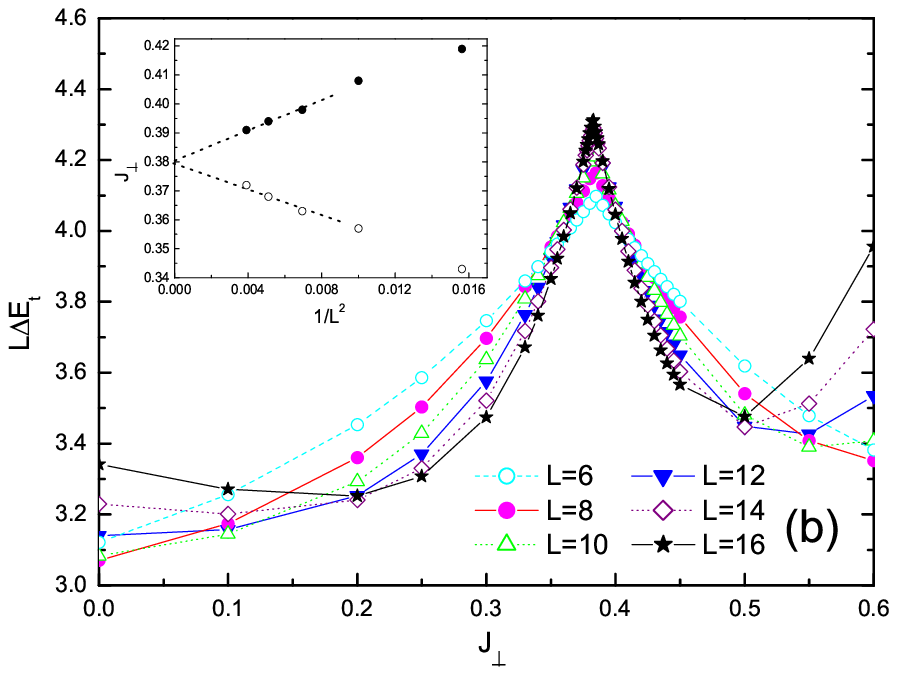} \vskip
-0.5cm \caption{(Color online) Scaled gaps (a) $L\Delta E_{s}$ for
spin-singlet excitation and (b) $L\Delta E_{t}$ for spin-triplet
excitation as functions of $J_\perp$ for various sizes $L$ with
$J_\times =0.2$. Insets show the $L^{-2}$ scaling behavior of the
left and the right crossing points represented by open and solid
circles, respectively. Dotted lines are guides to the eye. They
are straight lines fitted to the last few points.}
\label{fig:gaps}
\end{figure}

Finally, we provide a further examination of the possibility of
another critical point $(J_{\perp})_{c,S}$ with vanishing gap in
the spin-singlet sector. According to the phenomenological
renormalization-group (PRG) method,~\cite{Room80} second-order
phase transition points can be determined by the crossing points
of the curves of the scaled gaps $L \Delta E_\nu$ at two
successive sizes $L$ and $L+2$, where $\Delta E_\nu$ denote the
excitation gaps in the spin-singlet and the spin-triplet sectors
for $\nu= s$ and $t$, respectively. Here the gaps are calculated
by using exact diagonalization method with periodic boundary
conditions up to $L=16$. In the present case, the ground state is
unique for any $J_{\times}$ and $J_{\perp}$, and it has total spin
$S=0$ and momentum $k=0$. The spin-triplet (-singlet) excitation
gap $\Delta E_{t}$ ($\Delta E_{s}$) is determined by the energy
difference between the ground state and the lowest level with
total spin $S=1$ (with total spin $S=0$ and $k=\pi$). Our results
for the scaled gaps $L\Delta E_\nu$ as functions of $J_\perp$ for
various sizes $L$ with $J_\times =0.2$ are exhibited in
Fig.~\ref{fig:gaps}. For both cases of $L\Delta E_s$ and $L\Delta
E_t$, there are two crossing points of the curves at subsequent
sizes. However, it implies only one, but not two, critical point,
because the extrapolation of the left and the right crossing
points tend to a single value in the thermodynamical limit as
shown in the insets of Fig.~\ref{fig:gaps}. The limiting values
for the spin-singlet and the spin-triplet sectors are almost the
same, and both give $(J_\perp)_{c}\simeq 0.38$. Our findings
indicate that only one phase transition at $(J_\perp)_{c}\simeq
0.38$ occurs in the present system, and then the spin-singlet and
the spin-triplet gaps vanish simultaneously. We note that simple
linear extrapolations from data of systems of small sizes may be
somewhat dangerous. Nevertheless, because our results based on PRG
are consistent with that obtained by the above DMRG analysis and
that given by earlier DMRG calculation,~\cite{Wang} they may lead
to true physics in the thermodynamical limit.


In summary, we study numerically a two-leg spin ladder with
diagonal frustrations for weak interchain couplings. All our
results indicate that the transition from the RS to the Haldane
phases is direct, without any phase in between. This conclusion is
consistent with the picture obtained by all previous
investigations,~\cite{Gelfand,AEN,KFSS,FLS,Nakamura03,WEI,Wang,Nakamura-un}
except that proposed in Ref.~\onlinecite{Oleg} based on a
reformulated weak-coupling field theory. It is not clear why a
reformulated bosonization analysis works for one-dimensional EHM,
but could fail for the present two-leg spin ladder. Further
theoretical investigations are necessary to clarify this issue.

We thank M. Nakamura for very fruitful discussion in the early
stage. HHH and CDG are grateful for the support from the Ministry
of Science and Technology of China under Grant No.
NKBRSF-G19990646. YCC and MFY acknowledge the support by the
National Science Council of Taiwan under NSC 94-2112-M-029-003 and
NSC 94-2112-M-029-008, respectively.


\begin{thebibliography}{9}
\bibitem{note}
We also perform DMRG calculations for the case of $J_\times = 0.1$
and $J_\perp$ around $2J_\times = 0.2$. Same conclusion is
reached.
\bibitem{comment}
It is known that these string order parameters are not sufficient
to characterize uniquely the RS and the DC phases, because both
phases have even topological quantum number (i.e., ${\cal O}_{\rm
odd}=0$ but ${\cal O}_{\rm even}\neq 0$).~\cite{FLS} However,
since the possibility of the DC phase is ruled out by our results
in Figs.~\ref{fig:dimer} and \ref{fig:rung}, the only possible
phase with even topological quantum number is just the RS phase.
One then can use string order parameters to locate the regions of
the Haldane and the RS phases.
\end{thebibliography}
\end{document}